\documentclass[aps,prl,
noshowpacs,
10pt]{revtex4-2}
\usepackage[utf8]{inputenc}
\usepackage[T1]{fontenc}
\usepackage{graphicx}
\usepackage{dcolumn}
\usepackage{amsmath}
\usepackage{bm}
\usepackage{xcolor}
\usepackage{soul}
\usepackage{caption}
\usepackage{lineno}
\usepackage{lipsum}

\newcommand{\fa}{{(a)}}
\newcommand{\fb}{{(b)}}
\newcommand{\fc}{{(c)}}
\newcommand{\fd}{{(d)}}
\newcommand{\fe}{{(e)}}
\newcommand{\ff}{{(f)}}
\newcommand{\rfa}{(a)}
\newcommand{\rfb}{{(b)}}
\newcommand{\rfc}{{(c)}}
\newcommand{\rfd}{{(d)}}
\newcommand{\rfe}{{(e)}}
\newcommand{\rff}{{(f)}}
\newcommand{\figtit}{}
\newcommand{\methsec}{Methods}
\newcommand{\si}{Supplemental Material}

\begin{document}
    
\preprint{}


\title{Symmetry-group-protected microfluidics for multiplexed stress-free manipulations}

\author{Jeremias Gonzalez}
\affiliation{Department of Physics, University of California, Merced, Merced, CA 95343, USA}
\author{Ajay Gopinathan}
\affiliation{Department of Physics, University of California, Merced, Merced, CA 95343, USA}
\author{Bin Liu}
\affiliation{Department of Physics, University of California, Merced, Merced, CA 95343, USA}


\begin{abstract}

Modern micromanipulation techniques typically involve trapping using electromagnetic, acoustic or flow fields that produce stresses on the trapped particles thereby precluding stress-free manipulations. Here, we show that by employing polyhedral symmetries in a multichannel microfluidic design, we can separate the tasks of displacing and trapping a particle into two distinct sets of flow operations, each characterized and protected by their unique groups of symmetries. By combining only the displacing uniform flow modes to entrain and move targeted particles in arbitrary directions, we were able to realize symmetry-protected, stress-free micromanipulation in 3D. Furthermore, we engineered complex, microscale paths by programming and controlling the flow within each channel in real-time, resulting in multiple particles simultaneously following desired paths in the absence of any supervision or feedback. Our work therefore provides a general symmetry-group-based framework for understanding and engineering microfluidics and a novel platform for 3D stress-free manipulations.

\end{abstract}
\maketitle

\section{Introduction}

Microfluidics enables the study of the behavior, control, and manipulation of microscale flows typically using device architectures comprising channels and junctions as their building blocks \cite{squires_microfluidics_2005, whitesides_origins_2006, mark_microfluidic_2010,  battat_nonlinear_2022}. Desired flow structures and microfluidic functions such as transporting and mixing chemicals \cite{manz_planar_1992}, shaping flow profiles \cite{maoMicrofluidicDriftingImplementing2007, drescherBiofilmStreamersCause2013}, encapsulating multiphase fluids \cite{garstecki_formation_2004, amstad_microfluidic_2014}, and trapping microparticles \cite{perkins_single_1997, shenoy_stokes_2016} are incorporated through the geometry of these building blocks and the topology of their assemblies \cite{kaigala_microfluidics_2012, bhargavaDiscreteElements3D2014, au_3d-printed_2016, takahashi_stainless_2021}. 
Despite this increased complexity, our understanding of these microfluidic functions is mostly achieved in a case-by-case fashion relying on actual experimental measurements or on computational fluid dynamics (CFD) studies using detailed experimental geometries. A unified framework that allows us to understand how these complex flow structures emerge from simpler, more fundamental flows would prove extremely useful for the fast, modular development of rationally designed flow structures and associated microfluidic functions.

Symmetry is a concept that has been used to understand phenomena and structures in disciplines ranging from fundamental physics to viral structure \cite{coxeter_regular_1973, gross_physics_1996, zandi_origin_2004}. In the realm of microscale flows, symmetries are essential features of representative flow structures, such as the helical symmetry in swimming microorganisms \cite{purcellLifeLowReynolds1977, lauga_hydrodynamics_2009} and the reflection symmetry applicable in many hydrodynamic interactions \cite{berke_hydrodynamic_2008, elfring_hydrodynamic_2009}. These symmetries have profound implications with helical symmetry giving rise to swimming motility in bacterial species due to the breaking of kinematic reversibility while the reflection symmetries possessed by a pair of microswimmers constrain their synchronization and thus their collective motion. Such considerations applied to microfluidic systems could allow us to uncover very general results independent of system details. 
Recent work on multi-channel microfluidic junctions and open-space microfluidics have shown that their main flow structures can be obtained from the potential flows around charges representing the channels' ports, without consideration of detailed channel geometries \cite{shenoy_stokes_2016, goyette_microfluidic_2019}. 
This simplification has revealed intriguing flow structures that enable rich microfluidic applications, such as flow-driven traps for particle manipulation and dynamic confinements of fluid into multipole-like networks. 
Additionally, certain combinations of rotation and reflection symmetries have been applied both analytically and numerically to a multi-port microfluidic junction to eliminate strain components at the junction center, giving rise to potentially perturbation-free micromanipulations \cite{gonzalez_symmetry-based_2020}. 
All the above implications of symmetries suggest the feasibility of using a more generalized symmetry-based approach, i.e., the symmetry group, for a more fundamental understanding of the origins of these symmetries and ways to manipulate them for specific functions. 

One of the most desirable of these microfluidic functions is manipulating flow structures that allow the displacement of individual microscale objects along arbitrary paths \cite{fang_datadriven_2022}. Typically, micromanipulation techniques involve trapping particles using electromagnetic, acoustic, or flow fields that produce restoring forces in the vicinity of the trap \cite{ashkin_observation_1986, wu_acoustical_1991, grier_optical_1997, luan2020}. For example, in cases where hydrodynamic flow alone is used for trapping, previous studies have shown that a Stokes trap can be achieved by managing the locations of hyperbolic points through multiple channels that intersect at a middle junction \cite{shenoy_stokes_2016}. The location of such a trap can be dynamically adjusted to realize the direct manipulation of trapped particles, which has also been extended to 3D flows very recently \cite{tu_3d_2023}. 

\begin{figure}[ht]
\captionsetup{justification=raggedright,
singlelinecheck=false
}
\centering
\includegraphics[width=0.95 \textwidth]{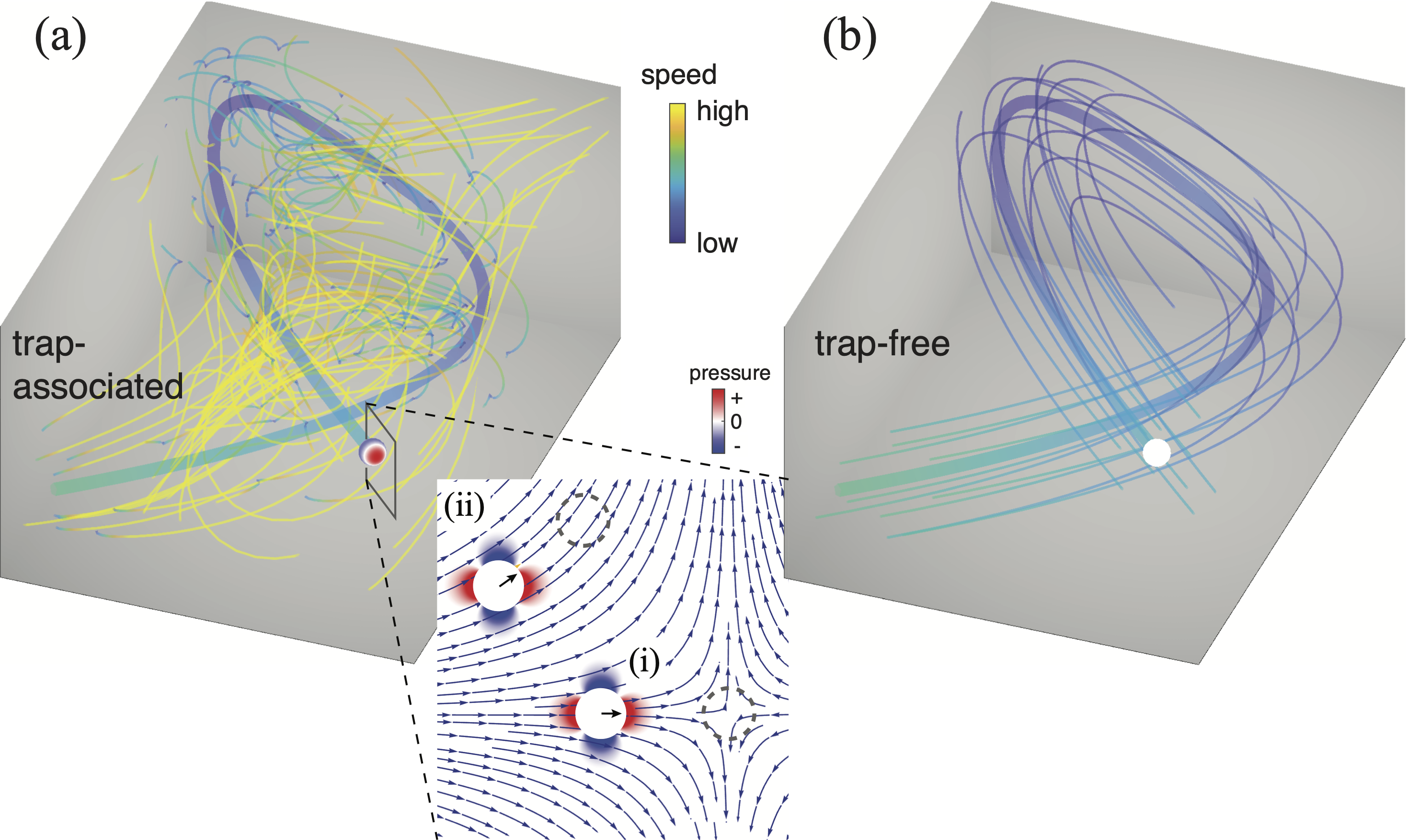}
    \caption{{\figtit Trap-associated v.s. trap-free manipulations of microscale particle through surrounding flows.} {\fa} In trap-associated manipulations, a manipulated particle is led by a fluid ``trap'' or guided by the local flow director (zoomed-in 2D views (i) and (ii) in the inset, respectively) along the desired pathway (shown in the thick curve), surrounded by overall divergent flows (with speeds of fluid tracers shown in colors) due to its trapping nature. The stress distribution around the particle (with dashed circles showing target positions) is intrinsic to the far-field flow topology (inset), with compression (in red) and extension (in blue) along orthogonal axes, independent of its relative position to the trap.  {\fb} An ideal perturbation-free manipulation is achieved by entraining the particle in a uniform flow field. All surrounding fluid flows in parallel paths at identical velocities and leaves the entrained particle stress-free (in blank color). }\label{fig:01}
\end{figure}

Computation-assisted feedback-control has also enabled manipulations of particles along local flow fields away from a hyperbolic point \cite{shenoy_flow_2019, fang_datadriven_2022}. The displacements of particles can be adjusted by aligning the local flow field with the moving directions of particles. However, in any such micromanipulation technique, trapping and displacement are inextricably coupled. As illustrated in Fig.~\ref{fig:01}{\rfa}, flow fields far from these manipulated particles are divergent, leading to inevitable stress perturbations on any manipulated objects. It is worth noting that the stress distribution on the particle is intrinsic to the far-field hyperbolic topology \cite{hyperbolicStress}. Even for a flow field that appears uniform away from the hyperbolic point (Fig.~\ref{fig:01}{\rfa} inset), the stress distribution on the entrained particle is identical to that within the trap, regardless of their relative displacement. These stressful perturbations thus restrict such approaches to studying stress-insensitive phenomena. Scaling such approaches with traps or local flow vector fields to simultaneously manipulate large numbers of particles is also a challenge. Here, we explore the use of symmetry groups to guide the design of microfluidics that enable multiplexed and stress-free manipulation, which is characterized by a truly uniform flow field around the manipulated particle(s) (Fig.~\ref{fig:01}{\rfb}). Specifically, we identify and characterize microfluidic flow structures based on symmetry groups, which allow us to distinguish and robustly realize fundamentally different microfluidic functions.

\section{Results}
\subsection{2D Microfluidic symmetries}
\begin{figure}[ht]
\captionsetup{justification=raggedright,
singlelinecheck=false
}
\centering
\includegraphics[width=0.95 \textwidth]{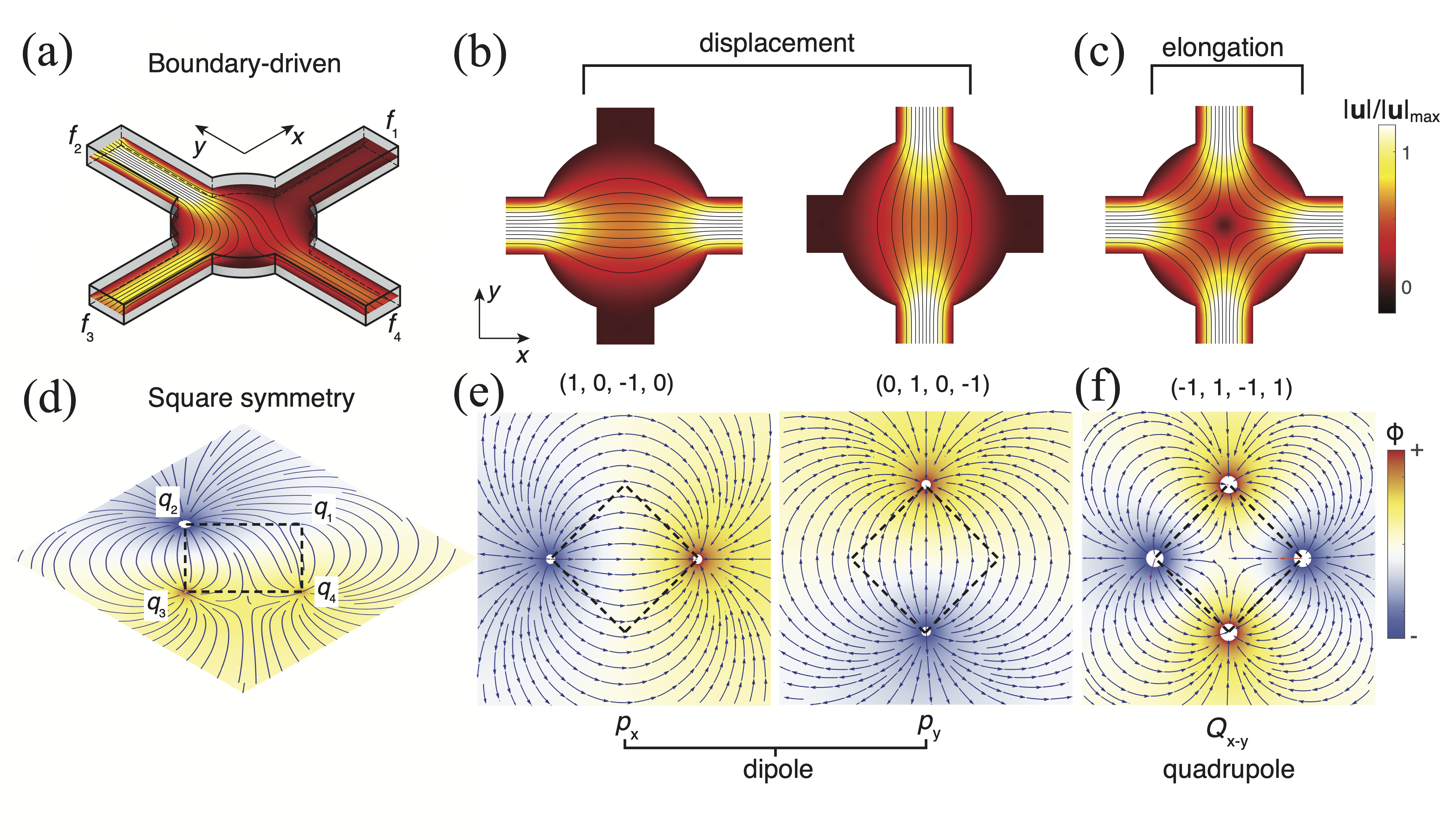}
\caption{{\figtit Functional classification of a cross-channel microfluidic junction. } {\fa} A 2D CFD simulation of the flow pattern (with streamlines in solid curves and speed $|\mathbf{u}|$ in colors) from an arbitrary flux setting ($f_1$ - $f_4$ at four ports) can be decomposed into three linearly independent flow modes: {\fb} two orthogonal displacement flow modes, and {\fc} one elongation flow mode. {\fd} The flow pattern in {\fa} is reproduced by a flow potential $\phi$ associated with fluid sources (or sinks) at the four corners of a square ($q_1$ - $q_4$). The square symmetry gives rise to {\fe} two degenerate dipole modes equivalent to two displacement modes in orthogonal directions, and {\ff} one quadrupole mode equivalent to elongation flow.}\label{fig:02}
\end{figure}

We start with a cross-channel junction for its well-understood microfluidic functions and relatively simple symmetries. Here, the four ports of identical channels (with flow rates $f_i$, $i=1,2,3,4$) form a square (Fig.~\ref{fig:02}{\rfa}), and therefore its flow structures in the plane are restricted by the symmetry of a square. An arbitrary flow pattern generated by a combination of $\{f_i\}$ at this junction corresponds to a set of identical flow manipulations through certain rotations or reflections ({\si}, Fig.~S1). All such possible rotations or reflections (including the trivial identity operation) form a dihedral-4 (D4) symmetry group \cite{coxeter_regular_1973}, which restricts the possible flow patterns. This D4 group thus represents the flow manipulation functions in a cross-channel junction. 

How do the D4 group elements map to the actual flow functions? Analogous to the subgroups contained in one symmetry group, these flow manipulation functions are reducible into smaller classes with unique features. We note first that there are only three independent flow modes, resulting from the number of controls (four channels) minus the number of constraints (one associated with the conservation of volume, i.e., $\sum_{i=1}^4 f_i=0$). Through inspection, one can readily identify these three modes, forming two classes of flow functions. First, fluxes with opposite signs through a pair of opposing channels give rise to displacement flows along two orthogonal directions ($x$ and $y$), corresponding to two fundamental modes (Fig.~\ref{fig:02}{\rfb}). Second, fluxes with same sign through both pairs of opposing channels provide one more fundamental elongation flow mode (Fig.~\ref{fig:02}{\rfc}). Any flow profile produced in this junction (with an example shown in Fig.~\ref{fig:02}{\rfa}) must be a combination of these three flow modes. However, none of these fundamental modes can be generated by combining any of the rest, a consequence of their linear independence. Importantly, the two displacement modes have a reflection symmetry (along $45^\circ$ from the $x$ axis), forming a reflection group D1 or a cyclic group C2, which is known to be a subgroup of the D4 group \cite{pal_physicists_2019}. The single elongation flow mode also represents a trivial subgroup of D4 that contains only the identity operation. 

In addition to forming different subgroups, these two classes of flows also have distinct characteristic functions, here displacement \textit{vs.} elongation. To abstract these functions from the detailed flow geometries, we map the cross-channel junction to a square, where the flux at each channel is mapped to an effective charge at each lattice site ($q_1$, $q_2$, $q_3$, $q_4$), representing a point-like fluid source (positive) or sink (negative). The flow field is thus well-defined as a potential flow, with the flow velocity $\mathbf{u}$ given by the gradient of the potential $\phi$, i.e., $\mathbf{u}=\boldsymbol{\nabla} \phi$. Since the classes of flow functions are independent of the choice of coordinates, the candidate quantities for such purposes are the strain rate invariants \cite{spencer_continuum_1980} that do not change under any rotation of the microfluidic device. The first-order invariant $I_1$, which is the trace of the rate of strain tensor $\dot{\gamma}$, is trivial and vanishes for incompressible fluids. We therefore consider the first non-trivial invariant $I_2=-\frac{1}{2}\mathrm{Tr}(\dot{\gamma}^2)$. For a 2D potential flow with a total number of $N=4$ sources and sinks (with magnitudes \{$q_i$\}, located at \{$\mathbf{r}_i$\}), the scalar potential can be expressed as $\phi(\mathbf{r})=\sum_{i=1}^4 q_i\ln\left(\left|\mathbf{r}-\mathbf{r_i}\right|\right)$. Since the rate of strain is linearly related to the magnitudes of sources (or sinks), we can show ({\si}, Equations S8-9) that the invariant $I_2$ at the center of the junction has a quadratic form with respect to $q_i$, i.e., $I_2=\sum_{i,j=1}^{4}L_{ij} q_i q_j$, where $L_{ij}$ is the element of a coupling tensor, depending on the distance between the pair of $q_i$ and $q_j$. On a square, there are only two possible types of pairs: pairs along the lateral sides or along the diagonals. With these considerations, we can show ({\si}, Equation S15) that the $\mathbf{L}$ tensor has the form
\begin{equation}
    \mathbf{L}=\left[
\begin{array}{cccc}
0 & a & b & a\\
a & 0 & a & b\\
b & a & 0 & a\\
a & b & a & 0
\end{array}\right].
\end{equation}

Elements ``a'' and ``b'' are symbolic representations of the above lateral and diagonal couplings, respectively, forming a symmetric circulant matrix for $\mathbf{L}$. A circulant matrix has all the same rows with each row being rotated one element to the right relative to the row above. One mutual eigenvector for any circulant matrix corresponds to the trivial mode where all four charges are identical, i.e., $(q_1,q_2,q_3, q_4)$ = $(1,1,1,1)$. This mode cannot be accessed by a fluid system due to the invalidation of the continuity condition. This leaves $N-1=3$ nontrivial eigenmodes (all satisfying $\sum_{i=1}^4 q_i=0$), in agreement with the degrees-of-freedom argument. Every nontrivial mode exhibits a one-to-one map to the fluidic manipulation function as previously classified through subgroups of the D4 group (Fig.~\ref{fig:02}{\rfd}). More specifically, the degeneracy of two orthogonal dipoles corresponds to a common displacement function of the flow along orthogonal directions (Fig.~\ref{fig:02}{\rfe}). The remaining quadrupole-like mode is nondegenerate and maps to the elongation function (Fig.~\ref{fig:02}{\rff}).   

\subsection{3D Microfluidic symmetries}

As demonstrated above, the eigenvalue analysis of the stress invariant ($I_2$) provides an automatic strategy of classifying microfluidic functions, which we now extend to 3D structures to incorporate more sophisticated symmetries. The potential flow now adopts its 3D form with $N$ sources, $\phi(\mathbf{r}) = \sum_{i=1}^N -\frac{q_i}{\left|\mathbf{r}-\mathbf{r}_i\right|}$. 

For relatively more practical applications in microfluidics, we consider here the 3D symmetries contained in structures with fewer vertices, namely the tetrahedron ($N=4$) and the octahedron ($N=6$). 

\begin{figure}[ht]
\captionsetup{justification=raggedright,
singlelinecheck=false
}
\centering
\includegraphics[width=0.95 \textwidth]{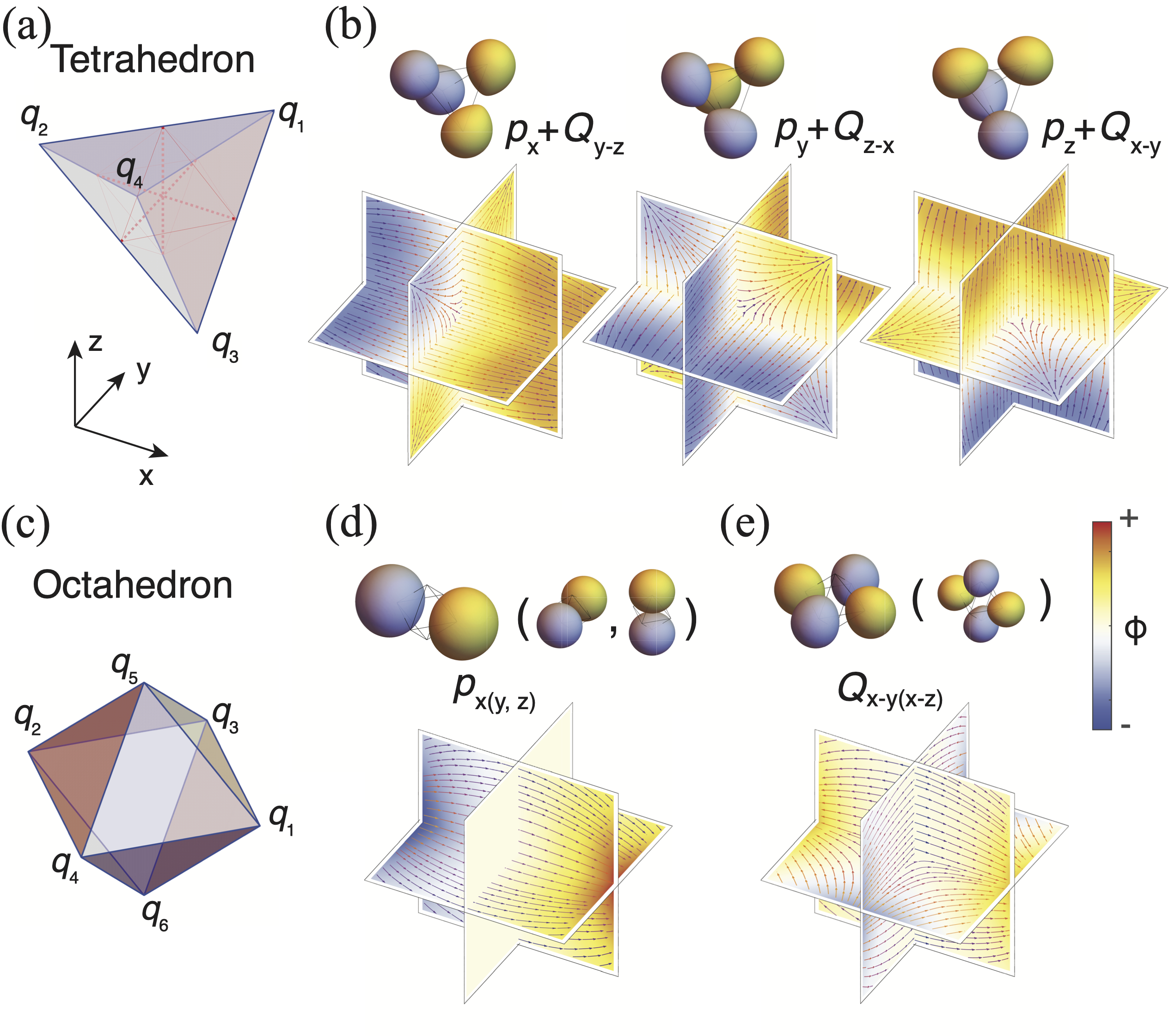}
\caption{{\figtit Microfluidic devices with 3D symmetries.} {\fa} Fluid sources (or sinks) following a tetrahedral symmetry ($q_1$ - $q_4$) can be classified into three degenerate functions, {\fb} a mix of dipole $p$ and quadrupole $Q$ in each of the three orthogonal directions, representing simultaneous displacement and elongation of the fluid. {\fc} Fluid sources (or sinks) following an octahedral symmetry ($q_1$ - $q_6$) can be classified into two subgroups: {\fd} three dipoles along orthogonal directions ($x$, $y$ and $z$) and {\fe} two quadrupoles in orthogonal planes (e.g., $x$-$y$ and $x$-$z$), contributing to separate displacement and elongation functions.}\label{fig:03}
\end{figure}

A tetrahedron has a permutation S4 symmetry \cite{pal_physicists_2019}, corresponding to an unaltered structure (identical through rotation and reflection) by permuting all four vertices (Fig.~\ref{fig:03}{\rfa}). The eigenvalue analysis shows that all three eigenmodes for this tetrahedron are degenerate (Fig.~\ref{fig:03}{\rfb}), due to equivalent neighbors for every vertex. These three modes together represent a threefold rotation symmetry group (C3) (achieved by $120^\circ$ rotation about any of the face norms), which is a subgroup of S4. Notably, the potential flow of such a mode exhibits a mixture of dipole and quadrupole moments (upper Fig.~\ref{fig:03}{\rfb}), leading to a microfluidic function corresponding to simultaneously displacing and elongating (transverse to the displacement axis) the fluid within the junction. 
To obtain a purely displacing flow in 3D, we therefore look to the next available polyhedral symmetry, i.e., the octahedral symmetry. 

An octahedron (Fig.~\ref{fig:03}{\rfc}) can be formed by the middle points of the six edges of a tetrahedron (Fig.~\ref{fig:03}{\rfa}). Its symmetry group is isomorphic to S4$\times$C2, with the additional symmetry arising from an extra 2-fold rotation symmetry that is absent for a tetrahedron (Fig.~\ref{fig:03}{\rfa}) \cite{pal_physicists_2019}. A similar eigenvalue analysis leads to two groups of degenerate modes, corresponding respectively to dipoles and quadrupoles. The three degenerate dipole modes ($p_x$, $p_y$, $p_z$) are orthogonal in direction, corresponding to a C3 group (through $120^\circ$ rotation about any of the face norms), one subgroup of the octahedral symmetry group \cite{pal_physicists_2019}. In such dipole modes, a pair of effective charges along the diagonal (e.g., $q_1$ and $q_2$ for $p_x$) have opposite signs (with the rest of the charges being neutral), corresponding to the activation of a pair of a source and a sink. The two degenerate quadrupoles are distributed in two perpendicular planes (here, $x$-$y$ and $x$-$z$ planes), forming a D1 subgroup (switchable by a reflection along the middle plane between $x$-$y$ and $x$-$z$ plane). In such quadrupole modes, a pair of sources and a pair of sinks along two diagonals are activated. It should be noted that an equivalent quadrupole in the $y$-$z$ plane can be achieved by superimposing the above two quadrupoles, and is thus not one of the eigenmodes. Analogous to the square symmetry in the 2D case, the octahedral symmetry gives rise to completely separate displacement and elongation flow functions (Fig.~\ref{fig:03}{\rfd} and {\rfe}), classified by distinct symmetries. The pure-dipole-like potential in the displacements also ensures that the rate-of-strain invariant $I_2$ is zero in the middle of the junction. Thus, a superposition of these three displacement flow modes generates 3D omnidirectional flows while preserving the stress-free condition (at least in the middle of the junction). 

\subsection{Realizing 3D Stress-free microfluidic manipulations}

\begin{figure}[ht]
\captionsetup{justification=raggedright,
singlelinecheck=false
}
\centering
\includegraphics[width=0.95 \textwidth]{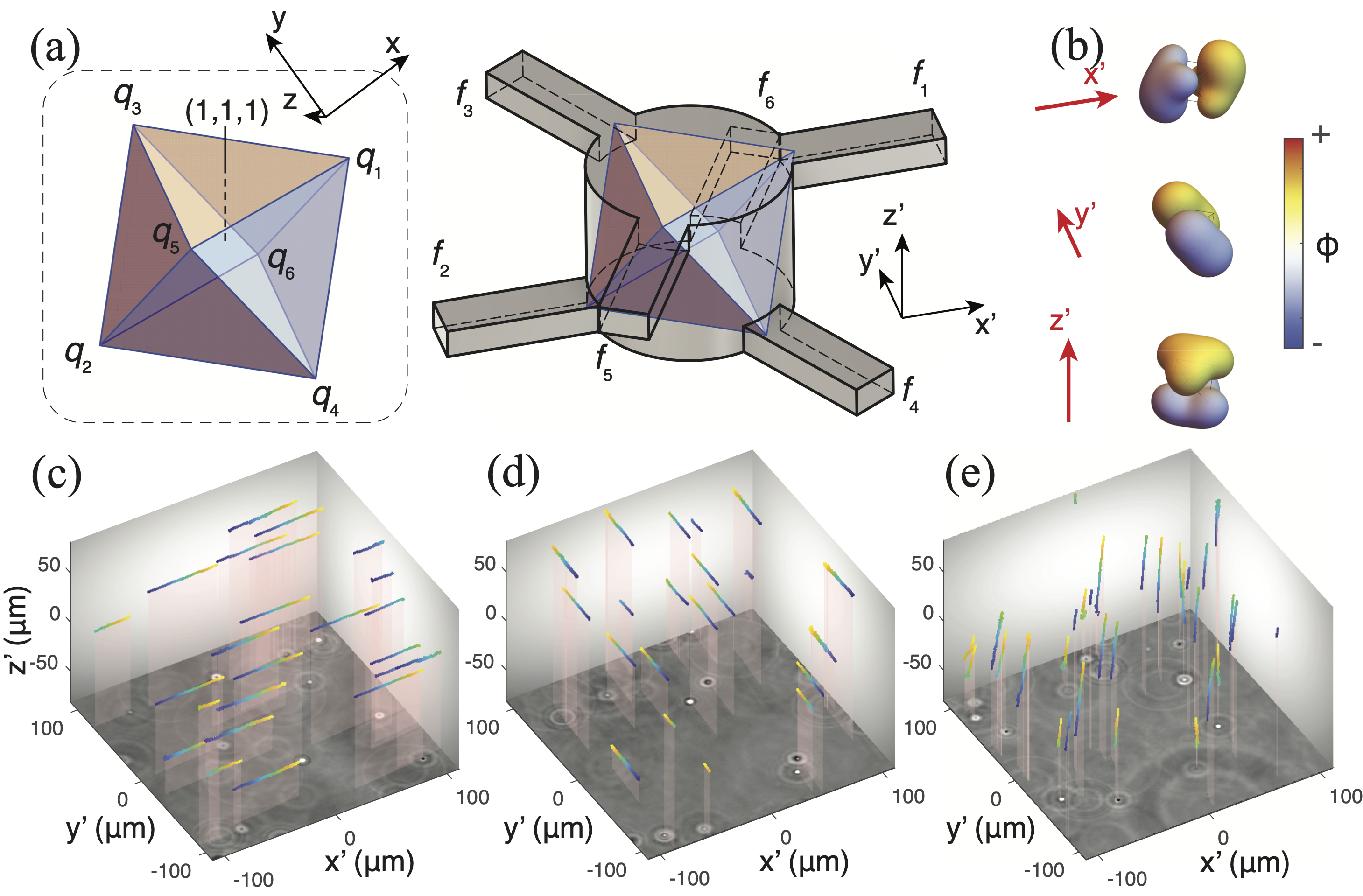}
\caption{{\figtit Realization of the octahedral symmetry in 3D microfluidics. } {\fa} An octahedron viewed along its original $(1,1,1)$ axis (inset) is converted into a microfluidic junction with all sources (or sinks) distributed at two planes perpendicular to the imaging axis $z'$. {\fb} The dipole flows in microscope coordinates $(x', y', z')$ are obtained by linear combination of the three orthogonal dipole modes (Fig.~\ref{fig:03}{\fd}), with their experimental realizations shown in \fc, \fd, and \fe, respectively. The axial $z'$ positions of the seeding particles are obtained by correlating their phase-contrast images (projected in the bottoms of volumes) with a calibrated library of diffraction patterns. }\label{fig:04}
\end{figure}

To realize such a stress-free microfluidic junction while enabling microscope observation, we rotate the octahedron so that one of its faces (e.g., $(1,1,1)$ axis) is aligned along the visualization axis ($z'$). This rotated geometry fits well into a double-layer microfluidic design \cite{gonzalez_symmetry-based_2020}: six channels ($0.5$ mm wide and $0.3$ mm deep in cross-sections) at two elevations intersecting a middle cylindrical chamber (with radius $R=1.0$ mm and height $H=1.6$ mm) along the radial directions, with the locations of intersections matching that of the octahedral vertices (Fig.~\ref{fig:04}{\rfa}). This rotation is also determined by the coordinates of the laboratory frame of reference $(x',y', z')$, with a radial channel (here, $f_1$) aligned along the $x'$ axis. We fabricated the microfluidic channel part of this device by fusing multiple sheets of laser-etched glass (Citrogene, see {\methsec}), which was then mounted on a customized adapter for flow control and microscopy. 

To ensure that a displacement flow can be robustly created in all possible directions, we first considered flows along the three orthogonal axes ($x'$, $y'$, $z'$) of the laboratory coordinates forming the basis of the velocity space. The potential representations (isosurfaces) of these orthogonal flows (along $x'$, $y'$, and $z'$) are shown in Fig.~\ref{fig:04}{\rfb}, which are linear combinations of the above three degenerate dipole modes with their coefficients $(c_1, c_2, c_3)$ being $\frac{1}{\sqrt{6}}(2, -1, -1)$, $\frac{1}{\sqrt{2}}(0, 1, -1)$, and $\frac{1}{\sqrt{3}}(1,1,1)$, respectively. In experiments, these effective charges are realized by independently offsetting the fluid pressure on the corresponding channel using two 4-channel piezoelectric regulators (Elveflow OB1 MK3+), with a positive offset for a positive charge (or source) and a negative offset for a negative one (or sink). Custom software was programmed to convert the above flow coefficients into actual pressure settings for flow generation ({\methsec}). The three orthogonal flows ($x'$, $y'$, $z'$) were successfully generated in our microfluidic device, within the same volume near the center (Fig.~\ref{fig:04}{\rfc}-{\rfe}), as demonstrated by the 3D traces of the seeding particles ({\methsec}). Despite their different geometries in flow dipoles (Fig.~\ref{fig:04}{\rfb}), these flows are all uniform over almost the entire 3D volume ($\approx$ $200$ $\mu$m $\times$ $200$ $\mu$m $\times$ $150$ $\mu$m) captured by the microscope, demonstrating the experimental realization of a robust ``stressless'' condition that is guaranteed by symmetry. 

\begin{figure}[ht]
\captionsetup{justification=raggedright,
singlelinecheck=false
}
\centering
\includegraphics[width=0.95 \textwidth]{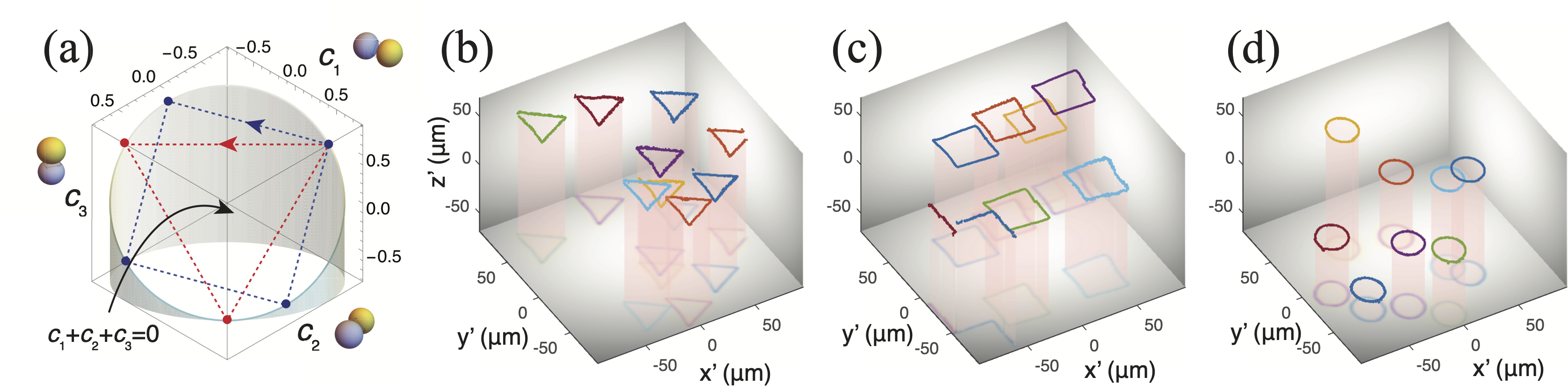}
\caption{{\figtit Dynamic stress-free displacement flows in 2D.} {\fa} Stress-free displacement flows were controlled in a phase space $(c_1, c_2, c_3)$, as the components of three orthogonal dipole modes. Traces in the $c_1+c_2+c_3=0$ plane of this phase space correspond to in-plane flows (in the $x'$-$y'$ plane) in the real space. The dashed lines correspond to instantaneous transitions between discrete states (dots) of the flow with the arrow showing the transition directions, forming a triangle, a square, and a circle in the continuous limit. The corresponding motions of the flow are shown in {\fb}, {\fc}, and {\fd}, respectively, with individual seeding particles shown in different colors. The periods of these patterns are $3$ s, $4$ s, and $4$ s, respectively.}\label{fig:05}
\end{figure}

To further investigate the experimental capacity of such stress-free microfluidics, we incorporated temporal dependence in the flow by dynamically varying the pressures on all channels. To facilitate this time-dependent control, we represented each possible configuration of the displacement flow by its coefficients $(c_1, c_2, c_3)$ when expressed in terms of three degenerate dipole modes, which essentially form a 3D phase space (Fig.~\ref{fig:05}{\rfa}). Any time-dependent flow manipulation can thus be generated by a series of points in this phase space. Noting that an axial ($z'$) flow is represented by a vector along the $\frac{1}{\sqrt{3}}(1,1,1)$ direction in the phase space, all orthogonal flows must satisfy $c_1+c_2+c_3=0$, which restricts all horizontal (or in-plane) displacement flows to a plane in the phase space. Hopping in this plane with equal angular separations and distances with respect to the origin $(0,0,0)$ gives rise to polygon-shaped flow patterns, for instance, a triangle, a square and a circle in the continuum limit (Fig.~\ref{fig:05}{\rfb}-{\rfd}, Movies S1-S3). Such patterns are almost identical for all seeding particles, with the fluid in the bulk translating like a piece of solid. 

\begin{figure}[ht]
\captionsetup{justification=raggedright,
singlelinecheck=false
}
\centering
\includegraphics[width=0.75 \textwidth]{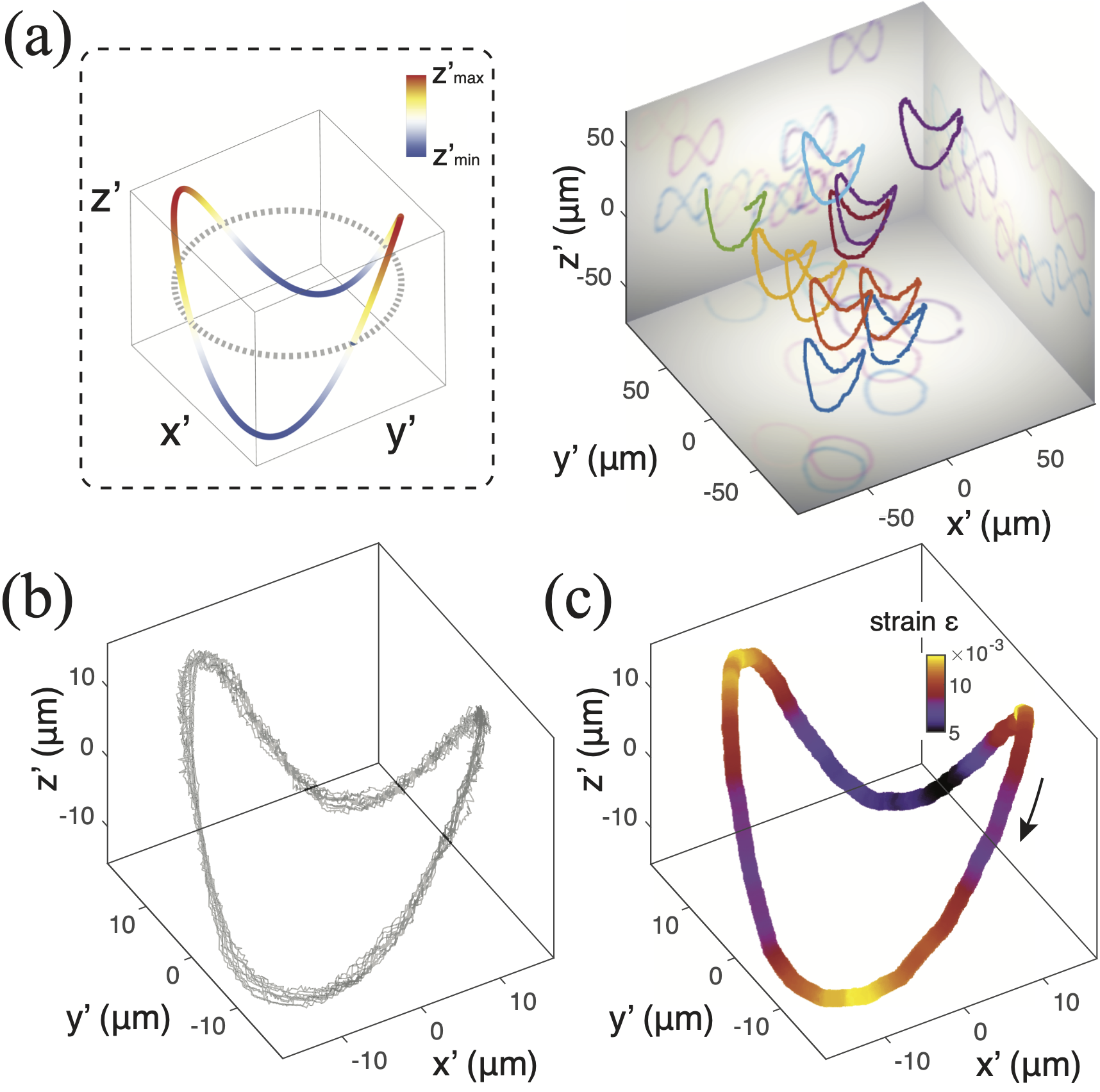}
\caption{{\figtit 3D stress-free and multiplexed manipulation.} {\fa} A design of the 3D Lissajous pattern (inset, with 3D positions in color and 2D projections in gray) was realized in the flow by incorporating additional axial motion into the circular pattern [Fig.~\ref{fig:05}{\rfd}], with all seeding particles in the view tracing out the desired 3D pattern (in $8$ s). (b) Visualized trajectories (gray curves) of seeding particles (11 particles shown) within the 3D view (200 $\mu m$ $\times$ 200 $\mu m$ $\times$ 150 $\mu m$ in $x' \times y' \times z'$) were overlaid and exhibited a fair degree of uniformity. (c) A characteristic strain $\epsilon$ computed along the manipulation path further confirms the uniformity (with $\epsilon$ bounded by $2\%$) of the manipulation flows. }\label{fig:06}
\end{figure}

By incorporating phase spaces out of the $c_1+c_2+c_3=0$ plane, we also generated time-dependent flows in 3D. For instance, imposing an oscillatory $z'$ motion (with doubled frequency) to the above circular mode gives rise to a 3D flow that represents a Lissajous curve (Fig.~\ref{fig:06}{\rfa}, Movie S4). Invariably, individual seeding particles within the volume of observation trace out the same pattern as desired.  demonstrating the robustness of the stressless condition even under 3D dynamic control. 

To assess the flow uniformity, we center the trajectory of every seeding particle within the entire volume and show them side by side. It is clear that these trajectories overlap with each other very well over a full period. In perfectly stress-free flows, all trajectories must be identical, leaving no deviation from the mean path. We thus defined a characteristic strain based on the deviation from the mean trajectory, i.e., $\epsilon (t) = \textrm{deformation}/\textrm{size} = \Delta \tilde{r}(t)/\Delta {r}(t)$, where $\Delta {r}\equiv\sqrt{\left<(\mathbf{r}-\left<\mathbf{r}\right>) \cdot (\mathbf{r}-\left<\mathbf{r}\right>) \right>}$ and $\Delta \tilde{r}(t)\equiv\sqrt{\left<(\tilde{\mathbf{r}}-\left<\tilde{\mathbf{r}}\right>) \cdot (\tilde{\mathbf{r}}-\left<\tilde{\mathbf{r}}\right>) \right>}$ are the fluctuations of displacements of all particles ($\mathbf{r}$) before (Fig.~\ref{fig:06}\rfa) and that ($\tilde{\mathbf{r}}$) after (Fig.~\ref{fig:06}\rfb) centering, with operators $\left<\cdot\right>$ corresponding to spatial averages. As shown in Fig.~\ref{fig:06}\rfc, such characteristic strains are of the order of $\epsilon \sim 10^{-2}$ for typical 3D manipulations, illustrating the nearly stress-free condition. 

\begin{figure}[ht]
\captionsetup{justification=raggedright,
singlelinecheck=false
}
\centering
\includegraphics[width=0.75 \textwidth]{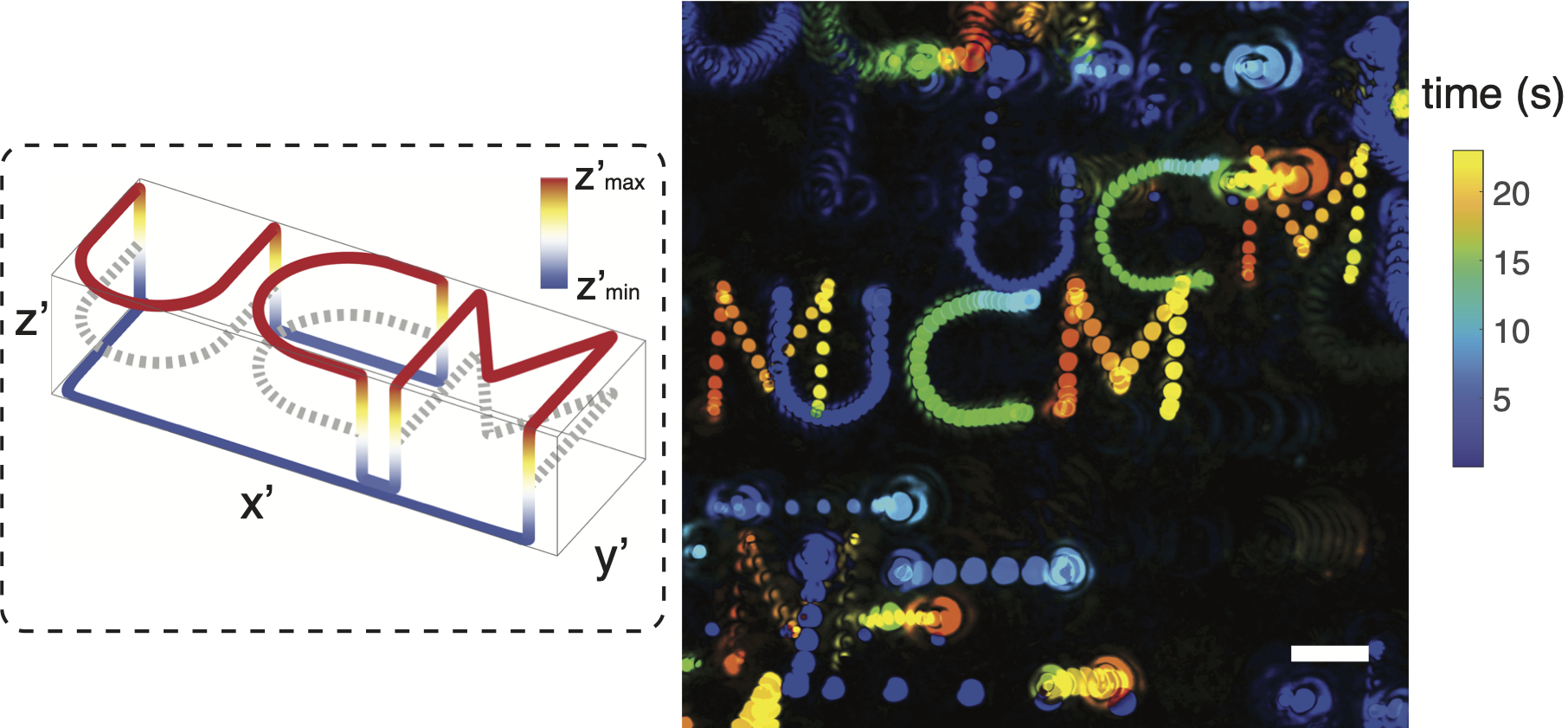}
\caption{More complex 3D flow manipulations were demonstrated by rapidly imprinting the ``UCM'' (with its design shown in the inset) onto the focal plane by seeding particles, which are color-coded in time (over $23$ s). Some of those particles initially out of the focal plane imprint the connecting lines (blue segments in the inset) between letters. The scale bar shows 20 $\mu m$.}\label{fig:07}
\end{figure}

The agile flow responses to the controlling pattern in the phase space indicate a direct map of the phase space to the real velocity space. The previously mentioned control in phase space can thus be generalized to arbitrary flow motions to realize more sophisticated manipulation capabilities. To demonstrate this concept, we combine both continuous motions and discrete hops in the phase space (or the velocities) to manipulate, in a stress-free manner, individual particles for ``printing'' discrete letters (here, ``UCM'') onto the focal plane. Each letter was traced out by combining motions (along smooth curves) and hops (at the corners of the letter) in the phase space. The discrete gaps between adjacent letters were achieved by abruptly offsetting the fluid in the axial direction so that the previously focused particles were moved away from the focal plane before their reappearance for ``printing'' the next letter. As shown in Fig.~\ref{fig:07}, this ``printing'' task can indeed be achieved in our stress-free microfluidic channel, which further demonstrates its versatile manipulation capabilities (Movie S5). 

\section{Discussion}

Our results show a substantial role for symmetry groups in classifying microfluidic functions. These symmetry-group-protected functions are insensitive to detailed geometries, which we exploited to realize 3D stress-free flows in a microfluidic device. Such flows are qualitatively different from all available micromanipulation approaches, where physical traps of certain forms must be present, resulting in stresses on the manipulated objects.  Both the stress-free modes and the stressful modes can be characterized by distinct subgroups of a group of polyhedral symmetry (possessed by the microfluidic device), suggesting independent transport and trapping in microfluidic manipulations. 

From our experimental observations, these stress-free flows are demonstrated by extremely parallel trajectories of the manipulated particles, regardless of the detailed curvatures of the trajectories. This enables us to achieve truly multiplexed stress-free manipulations, a significant challenge for trap-based approaches. It is worth noting that our trajectories inevitably deviate from the ideal geometries due to the absence of feedback controls. Meanwhile, any potential differences (e.g., different flow resistances) among the six channels lead to channel-sensitive responses to applied pressures, which ultimately modifies our time-dependent flow patterns. Remarkably, these ``defects'' in manipulations do not alter the uniformity in the flows (with all particle trajectories remaining parallel), suggesting the robustness in the symmetry-protected flows.  

Our work therefore opens up new avenues of experiments on microscopic phenomena that occur in truly stress-free flows.
Combinations of the above subgroups easily lead to other subgroups with higher orders (e.g., the C4 subgroup of the D4 group), associated with flow characteristics more than the strain invariants at the center. Flow patterns belonging to these subgroups maybe coordinated in time to form a mutual impact on the manipulated flow, e.g., creating a ``mean'' vortex that is absent from steady potential flows. These advanced symmetry features and flow manipulation functions will be explored in our future work.  

\section*{Acknowledgements}

This work was supported by National Science Foundation Grant CBET-2046822, CBET-1706511, NSF-CREST: Center for Cellular and Bio-molecular Machines (CCBM) at UC Merced (HRD-1547848), and Department of Defense DURIP ARO No.73839-MA-RIP.

\section{Appendix: Methods}
\textbf{Microfluidic fabrication and its assembly for microscope observations.} The microfluidic channel was fabricated as a stack of glass sheets (0.13 - 1 mm thick), with each containing a special pattern of channels. We made the design of these sheets through a computer assisted design (CAD) software (Autodesk Fusion 360), which were then sent to Citrogene for fabrications of these customized sheets in borosilicate glass and their integration into one single multilayer microfluidic chip (using optically clear adhesives). The overall exterior dimension of the microfluidic chip is compact, here, 62 mm $\times$ 22 mm $\times$ 2.7 mm in $x$, $y$ and $z$, comparable to the footprint of a typical glass slide. The chip was also designed to be self-enclosed, leaving only 6 small pores (1 mm in diameter) open for accessing the microfluidic flows (see {\si}). A 0.13 mm thick glass sheet at the bottom layer sealed the microfluidic channels and enabled high optical quality for conventional microscope observations. Before being connected to a multi-channel pump, the glass chip was mounted to a customized adapter for better sealing results. 

\textbf{Multiple-channel flow controls.} Flows within the multilayer microfluidic channel were generated by two 4-channel microfluidic flow controllers (Elveflow OB1 MK3+) with each channel's pressure regulated independently between -0.9 and 1 bar relative to the atmospheric pressure. A custom program (written in Python) was used to modulate the pressures on each channel and record the microscope image in real time. To accommodate the finite resolution of our pressure controllers, here 120 $\mu$bar, we used a water-glycerol mixture to increase the viscosity of the fluid in the microfluidic channel, which maintained the flow speed within a reasonable range for microscope observations while using a decent fraction of the pressure range. 

\textbf{Microscope imaging and three-dimensional flow reconstructions.}
The microfluidic flows were visualized by mixing the fluid with polystyrene beads ($2$ $\mu$m in diameter) as seeding particles, imaged under an inverted microscope (Nikon Eclipse Ti2) at a $60\times$ magnification, operated in its phase-contrast mode. These phase-contrast images were recorded by a USB Scientific CMOS (sCMOS) video camera (Andor Zyla 4.2) in full resolution ($2048\times 2048$ pixels) at 50 fps. The diffraction pattern of each seeding particle was visible within a range of $160$ $\mu$m along the optical axis, which were calibrated to restore the axial positions of all seeding particles in the view \cite{taute_high-throughput_2015}, forming a sizable (at least $200$ $\mu$m $\times$ $200$ $\mu$m $\times$ $100$ $\mu$m) visible zone of the 3D flow field.

\newpage

\end{document}